\documentclass[12pt,twosided]{article}
\usepackage[centertags]{amsmath}
\usepackage{amsfonts}
\usepackage{amssymb}
\begin{document}
\title{Basic Twist Quantization of $osp(1|2)$ and $\kappa$--Deformation
of $D=1$ Superconformal Mechanics}

\author{A. Borowiec$^{1)}$, J. Lukierski$^{1), 2)}$ and V.N.
Tolstoy$^{1), 3)}$
\\ \\
$^{1)}$Institute for Theoretical Physics, \\
 University of Wroc{\l}aw,
pl. Maxa Borna 9,                             \\
 50--205 Wroc{\l}aw, Poland
\\ \\
$^{2)}$Departimento de Fisica Teorica,\\
 Universidad de Valencia,  Av. Dr. Moliner 50,\\
46100 Burjasot (Valencia), Spain
\\ \\
$^{3)}$       Institute of Nuclear Physics,
  \\
Moscow State University, 119992 Moscow, Russia}

\date{}
\maketitle \thispagestyle{empty}
\begin{abstract}
The twisting function describing a nonstandard (super-Jordanian)
quantum deformation of $osp(1|2)$ is given  in explicite closed
form. The quantum coproducts and universal $R$-matrix are
presented. The non-uniqueness of the twisting function as well as
two real forms of the deformed $osp(1|2)$ superalgebras are
considered. One real quantum $osp(1|2)$ superalgebra is
interpreted as describing the $\kappa$-deformation of $D=1,\, N=1$
superconformal algebra, which can be applied as a symmetry algebra
of $N=1$ superconformal mechanics.
\end{abstract}

\newpage
\setcounter{page}{1}

\section{Introduction}

It is well-known that for the Lie algebra $sl(2;\mathbb R)\simeq o(2,1)$
do exist only two inequivalent deformations, generated by the classical
$r$-matrices with the following antisymmetric parts
($a\wedge b\equiv a\otimes b-b\otimes a$):

i) Standard deformation \cite{PK,bor1,bor2,bor3} with
\begin{equation}\label{boro1.1}
r_{\!_{ DJ} }=\gamma\,e_{+}^{}\wedge e_{-}^{}\, ,
\end{equation}
where
\begin{equation}\label{boro1.2}
[h,\,e_{\pm}^{}]=\pm e_{\pm}^{}\,,\qquad [e_{+}^{},\,e_{-}^{}]=2h \,
\end{equation}
defines the Cartan-Chevalley basis of $sl(2;\mathbb R)\simeq
o(2,1)$, and $\gamma$ is a deformation parameter. Classical
$r$--matrix (\ref{boro1.1}) determines the term linear in
deformation parameter in the coproduct of Drinfeld-Jimbo quantum
algebra $U_q(sl(2))$ where $sl(2)= sl(2;\mathbb R)$ or
$sl(2;\mathbb C)$. The classical $r$-matrix (\ref{boro1.1})
satisfies the {\it modified} YB equation (MYBE).

ii) Nonstandard (Jordanian) deformation \cite{bor4,bor6,bor7} with
\begin{equation}\label{boro1.3}
r_{\!_{J}}^{}=\xi\,h\wedge e_{+}^{}\,.
\end{equation}
The classical ${r}$-matrix (\ref{boro1.3}) satisfies the {\it classical} YB
equation (CYBE), and the corresponding quantization of $U(sl(2))$ can be
obtain by using Drinfeld twist technique \cite{bor8,bor9} with the twisting
two-tensor \cite{bor6}
\renewcommand{\theequation}{4\alph{equation}}
\setcounter{equation}{0}
\begin{equation}\label{boro1.4}
F_{\!_{J}}=\exp(\xi\,h \otimes E_{+}^{})\, ,
\end{equation}
where
\begin{equation}\label{boro1.5}
E_{+}^{}=\frac{1}{\xi}\,\ln(1+\xi\,e_{+}^{})=e_{+}^{}+{\mathcal O}(\xi)
\end{equation} describes deformed the generator
$e_{+}^{}$ with proper no-deformation limit $\xi\to\infty$. The
quantum Hopf algebra $U_\xi (sl(2))$ has the classical
non-deformed  algebra sector but the deformed twisted coproduct and
antipode
\renewcommand{\theequation}{\arabic{equation}}
\setcounter{equation}{4}
\begin{equation}\label{boro1.6}
\Delta _\xi (a) = F_{\!_{J}}^{}\Delta^{(0)}(a)F^{-1}_{\!_{J}}\, ,
\qquad S_{\xi} (a) = u S(a) u^{-1} \, \qquad (a\in U(sl(2)))\, ,
\end{equation}
where $\Delta^{(0)}(a)=a\otimes 1 + 1 \otimes a$ for any $a\in sl(2)$,
and $u = \sum_i f^{(1)}_i S(f^{(2)}_i)$ provided that the twist factor
(\ref{boro1.4}) is written $ F_{_{J}}=\sum_i f^{(1)}_i \otimes f^{(2)}_i$.
The nonstandard quantum Hopf algebra $U_\xi (sl(2))$ can be also described
in a nonclassical algebra basis \cite{bor7} with deformed $sl(2)$
Lie-algebraic relations.

The  algebra (\ref{boro1.2}) with the following reality conditions
\begin{equation}\label{real}
e_{\pm}^\dag = - \,e_{\pm}^{}\,,\qquad h^{\dag}_{}=-h
\end{equation}
can be considered as $D=1$ conformal algebra $sl(2;R)\simeq o(2,1)$
\begin{equation}\label{boro1.7}
[{\cal D},\,{\cal H} ]=i\,{\cal H}\,,\qquad [{\cal D},\,{\cal K}]=
-i\,{\cal K}\,,\qquad [{\cal H},\,{\cal K}]=-2\,i\,{\cal D}\,,
\end{equation}
where ${\cal H}$ describes the time translation generator (Hamiltonian),
${\cal D}$ is a scale generator, and ${\cal K}$ is a conformal acceleration
generator. We can identify (\ref{boro1.2}) and (\ref{boro1.7}) if
\begin{equation}\label{boro1.8}
e_{+}^{}=i\,{\cal H}\,,\qquad e_{-}^{}=-i\,{\cal K}\,,\qquad h= - i\,{\cal D}\,.
\end{equation}
One gets
\begin{equation}\label{boro1.9}
r_{\!_{DJ}}=\gamma\,{\cal H}\wedge{\cal K}\,,\qquad
r_{\!_{J}}=\xi\,{\cal D}\wedge{\cal H}\,.
\end{equation}
Further we shall choose classical $r$-matrix purely imaginary
under the involution (\ref{real})\footnote{In such a case we
obtain a form of quantum group, which was also called an imaginary
form (see e.g. \cite{VL})}, what implies after the assumption
$(a\otimes b)^\dagger =a^\dagger\otimes b^\dagger$ that $\gamma$
and $\xi$ are purely imaginary. We see that under the physical
scaling transformations
\begin{equation}\label{boro1.10}
{\cal H}'=\lambda\,{\cal H} \,,\qquad{\cal K}'=\lambda^{-1}\,{\cal K}\,,
\qquad{\cal D}'={\cal D}\, ,
\end{equation}
the parameter $\gamma$ is dimensionless, and $\xi$ has the inverse
mass dimension. If we put $\xi = \frac{-i}{2\kappa}$ we see that
the Jordanian deformation of $sl(2)$ describes the
$\kappa$-deformation of $D=1$ conformal algebra \cite{bor11} with
$\kappa$ describing the fundamental mass deformation parameter and
classical limit given by $\kappa\to\infty$ ($\xi \to 0$).

The aim of this paper is to provide explicite formulae for the super-Jordanian
twist quantization of the superalgebra $osp(1|2;\mathbb R)$ which is
supersymmetric extension of $sl(2;\mathbb R)\simeq sp(2;\mathbb R)\simeq o(2,1)$
\cite{bor12,bor13}. The classical $r$-matrices (\ref{boro1.1}) and (\ref{boro1.3})
are supersymmetrically extended as follows \cite{bor14}
\renewcommand{\theequation}{11\alph{equation}}
\setcounter{equation}{0}
\begin{eqnarray}\label{boro1.11a}
r_{_{DJ}}^{susy}&=&\gamma (e_{+}^{}\wedge e_{-}^{}+2v_{+}^{}\wedge v_{-}^{})\,,
\\[9pt] \label{boro1.11b}
r_{_{J}}^{susy}&=& \xi(h\wedge e_{+}^{}-v_{+}^{}\wedge v_{+}^{})\,,
\end{eqnarray}
where for odd generators $a\wedge b=a\otimes b+b\otimes a$, and
the standard relations of the $osp(1|2)$ Cartan--Chevalley basis looks as
follows
\renewcommand{\theequation}{12\alph{equation}}
\setcounter{equation}{0}
\begin{eqnarray}\label{boro1.12a}
[h,\,v_{\pm}^{}]=\pm\,\frac{1}{2}\,v_{\pm}\,, \quad\!\!\!&&\!\!\!\quad
\{v_{+}^{},\,v_{-}^{}\}=-\frac{1}{2}\,h\,,
\\[7pt]
\label{boro1.12b} e_{\pm}^{}\!\!&=&\!\!\pm4\,(v_{\pm}^{})^2\,.
\end{eqnarray}
One sees that $e_{\pm}^{}$ play the role of composite "double
root" generators, extending $osp(1|2)$ Cartan--Chevalley basis
(\ref{boro1.12a}).
\renewcommand{\theequation}{\arabic{equation}}
\setcounter{equation}{12}

One can consider two different reality conditions which represent
two possible supersymmetric extensions of the antilinear
antiinvolution defining (\ref{real}) -- we shall call them {\it
Hermitean} and  {\it graded} or {\it super-Hermitean}:
\renewcommand{\theequation}{13\alph{equation}}
\setcounter{equation}{0}

$(i)$ The Hermitean  reality conditions defined as follows
\begin{eqnarray}
\label{boro1.13a} (v_\pm)^\dag\!\!&=&\!\! i\,v_\pm
\end{eqnarray}
provided that
\begin{eqnarray}
\label{boro1.13a'}(ab)^{\dag}=b^{\dag}a^{\dag} \ \ \ \ {\rm and}\
\ \  (a\otimes b)^{\dag}=(-1)^{\deg a \deg b}\,a^{\dag}\otimes
b^{\dag}\end{eqnarray} for any homogeneous elements $a,\,b\in
U(osp(1|2))$.

$(ii)$ The  super-Hermitean reality conditions we define
\begin{eqnarray}
\label{boro1.13b} (v_\pm)^\ddag\!\!&=&\!\!v_\pm
\end{eqnarray}
provided that
\begin{eqnarray}
\label{boro1.13b'}(ab)^{\ddag}=(-1)^{\deg a \deg
b}\,b^{\ddag}\,a^{\ddag} \ \ \ \ {\rm and}\ \ \  (a\otimes
b)^{\ddag}=\,a^{\ddag}\otimes b^{\ddag}\end{eqnarray} for any
homogeneous elements $a,b\in U(osp(1|2))$.
\renewcommand{\theequation}{\arabic{equation}}
\setcounter{equation}{13}

The reality conditions (\ref{boro1.13a}) and (\ref{boro1.13b}) for
odd generators will provide new real $osp(1|2)$ $\star$-Hopf
superalgebras generated by the classical $r$-matrix
(\ref{boro1.11b}).

The quantization with $r_{_{DJ}}^{susy}$ describes Drinfeld--Jimbo
type of the quantum superalgebra, $U_q(osp(1|2))$, studied firstly
in \cite{bor15,bor16}. The quantization with $r_{_{J}}^{susy}$ is
a subject of the present paper. One can show that
(\ref{boro1.11b}) satisfies graded CYBE and can be quantized by
superextension of the Drinfeld twisting procedure. Further
introducing the $D=1$ conformal supercharges $Q,\,S$ which
transform under scaling (\ref{boro1.10}) as follows
\begin{equation}\label{boro1.14}
  Q' = \lambda^{1/2} \, Q\, , \qquad
  S' = \lambda^{- 1/2} \, S \, ,
\end{equation}
we easily see that the classical $r$-matrix (\ref{boro1.11b})
describes via identification $\xi = \frac{1}{2\kappa\,i}$ the
$\kappa$-deformation of $D=1, N=1$ superconformal algebra.

We would like to point out that in this paper we complete the discussion of
the super-Jordanian deformation of $osp(1|2)$ presented in \cite{bor14,bor19}.
In \cite{bor19} the proposed ansatz for the super-Jordanian twisting two-tensor
was not properly chosen what subsequently did not allow to complete the twist
quantization procedure generated by the classical $r$-matrix (\ref{boro1.11b}),
and in particular it was not possible to write down  all coproduct formulae.
It should be admitted however, that several results on the super-Jordanian
deformation of $osp(1|2)$ which did not require the complete knowledge of
the twisting two-tensor were presented in \cite{bor14,bor19}.

The plan of our paper is the following:

In Sect. 2 we present an explicite formula for the super-Jordanian
twisting two-tensor, calculate coproducts for all generators and
present the universal $R$-matrix.
We shall also discuss in Sec. 2 the real forms of quantum
$osp(1|2)$ superalgebras as well as non-uniqueness of the twisting
procedure. In Sect. 3 we interpret the super-Jordanian deformation
of $osp(1|2;\mathbb R)$ generated by the classical $r$--matrix
(\ref{boro1.11b}) as a $\kappa$-deformation of the $D=1$
superconformal algebra. Final Sect. 4 contains an outlook.

\section{Super-Jordanian Twist Quantization of $osp(1|2)$}

Firstly we shall outline basic elements of Drinfeld`s theory of twisting
quantization of Hopf algebras \cite{bor9}. A Hopf algebra
$A:=A(m,\Delta,\epsilon,S$) with a multiplication
$m:A\otimes A\rightarrow A$, a coproduct $\Delta:A \rightarrow A\otimes A$,
a counit $\epsilon:A\rightarrow\mathbb C$, and an antipode $S:A\to A$ due to
twisting procedure can be transformed with a help of an invertible element
$F\in A\otimes A$, $F=\sum_i f^{(1)}_i\otimes f^{(2)}_i$, into a twisted Hopf
algebra $A_\xi:=A_\xi(m,\Delta_\xi,\epsilon,S_\xi$) which has the same
multiplication $m$ and the counit mapping $\epsilon$ but the twisted coproduct
and antipode
\begin{equation}\label{boro2.1}
\Delta_\xi (a)=F\Delta(a)F^{-1},\quad\;S_\xi(a)=u\,S(a)u^{-1},
\quad\;u= \sum_i f^{(1)}_{i}S(f^{(2)}_i)\,\quad\;
(a\in A)\,.\end{equation}
The twisting element (twisting two-tensor) $F$ satisfies the cocycle equation
\begin{equation}\label{boro2.2}
F^{12}(\Delta\otimes{\rm id})(F)=F^{23}({\rm id}\otimes\Delta)(F)\,,
\end{equation}
and the "unital" normalization condition
\begin{equation}\label{boro2.3}
(\epsilon \otimes{\rm id})(F)=({\rm id}\otimes\epsilon )(F)=1\,.
\end{equation}
The Hopf algebra $A$ is called quasitriangular if it has an
additional invertible element (universal $R$-matrix) $R$
\cite{bor1,bor9} which relates the coproduct $\Delta$ with its
opposite coproduct $\Delta^{^{\rm op}}$ by the similarity
transformation
\begin{equation}\label{boro2.4}
\Delta^{^{\rm op}}(a)=R\,\Delta(a)R^{-1}\,\qquad (a\in A)\,,
\end{equation}
with $R$ satisfying the quasitriangularity conditions
\begin{equation}\label{boro2.5}
(\Delta\otimes {\rm id})(R)=R^{13}R^{23}\,,\qquad
({\rm id}\otimes\Delta)(R)=R^{13}R^{12}\,.
\end{equation}
The twisted ("quantized") Hopf algebra ${\mathcal A}_\xi$ is also
quasitriangular with the universal $R$-matrix $R_\xi$ defined as
follows
\begin{equation}\label{boro2.6}
R_{\xi} = F^{21}R\,F^{-1}\,,
\end{equation}
where $F^{21}=\sum_i f^{(2)}_i\!\otimes f^{(1)}_{i}$ provided
$F=\sum_i f^{(1)}_i\!\otimes f^{(2)}_i$. For the nondeformed, classical case
$A=U(g)$, where $g$  is a simple Lie algebra, the universal $R$-matrix is
trivial, i.e. $R=1$.

Our goal is to construct the twisting two-tensor $F_{_{\!SJ}} :=
F$ for $U(osp(1|2))$, such that the universal $R$-matrix
\begin{equation}\label{boro22.6}
R_{_{\!SJ}}=F^{21}_{_{\!SJ}}F^{-1}_{_{\!SJ}}\,
\end{equation}
has the form
\begin{equation}\label{boro22.7}
R_{_{\!SJ}}^{}=1-r_{_{\!SJ}}+{\mathcal O}(\xi^2)\,,
\end{equation}
where $r_{_{\!SJ}} $ is the classical $r$-matrix (\ref{boro1.11b}) linear
in deformation parameter $\xi$. The Taylor-series expansion of $F_{_{\!SJ}}$
with respect to the parameter $\xi$ looks as follows
\begin{equation}\label{boro22.8}
F_{_{\!SJ}}=1+\xi(h\otimes e_{+}^{}-v_{+}^{}\otimes v_{+}^{})+{\mathcal O}(\xi^2)
\end{equation}
and it is consistent with the relations (\ref{boro22.6}), (\ref{boro22.7}).
One can show that (see also \cite{bor19}) the twisting two-tensor $F_{_{\!SJ}}$
describing the quantization of the classical $r$-matrix (\ref{boro1.11b})
can be factorized as follows:
\begin{equation}\label{boro221}
F_{_{\!SJ}}=F_{_{\!S}}\, F_{_{\!J}}\,,
\end{equation}
where $F_{_{\!J}}$ is the Jordanian twisting two-tensor (\ref{boro1.4}) for
$sl(2)$ \cite{bor6} depending on $h$ and $\sigma$ (i.e. $h$ and $e_{+}^{}$)
and $F_{_{S}}$, the supersymmetric part, depends only on odd generator
$v_{+}^{}$ (remember $e_{+}^{}=4v_{+}^{2}$). The Jordanian twisting two-tensor
$F_{_{\!J}}$ as well as $F_{_{\!SJ}}$ should satisfy the cocycle and "unital"
conditions (\ref{boro2.2}), (\ref{boro2.3}).

Substituting (\ref{boro221}) into (\ref{boro2.2}) one obtains the following
twisted cocycle condition for $F_{_{\!S}}$
\begin{equation}\label{boro22.4}
F^{12}_{_{S}}(\Delta_{_{\!J}}^{}\otimes1)(F_{_{\!S}}^{})=
F^{23}_{_{S}}(1\otimes\Delta_{_{\!J}}^{})(F_{_{\!S}}^{})\,,
\end{equation}
where
\begin{equation}\label{boro225}
\Delta
_{_{\!J}}(a)=F^{}_{_{J}}\Delta^{(0)}(a)F_{_{\!J}}^{-1}\qquad(a\in
U(sl(2)))\,.
\end{equation}
We mention that because $sl(2)\subset osp(1|2)$ the classical $r$-matrix
(\ref{boro1.3}) for $sl(2)$ satisfying CYBE can be considered also as the
classical $r$-matrix for $osp(1|2)$. The coproducts (\ref{boro225}) have
been calculated in \cite{bor18} and are given by the formulae:
\renewcommand{\theequation}{27\alph{equation}}
\setcounter{equation}{0}
\begin{eqnarray}\label{boro226a}
\Delta_{_{\!J}}(e^{\pm\sigma})\!\!&=&\!\!e^{\pm\sigma}\otimes
e^{\pm\sigma}\,,\\[3pt]\label{boro226a'} 
\Delta_{_{\!J}}(e_{+}^{})\!\!&=&\!\!\frac{1}{\xi}\,(e^{2\sigma}\otimes
e^{2\sigma}-1)=e_{+}^{}\otimes e^{2\sigma}+1\otimes e_{+}^{}\,,
\\[3pt]\label{boro226b}
\Delta_{_{\!J}}(h)\!\!&=&\!\!h\otimes e^{-2\sigma}+1\otimes
h\,,\\[9pt]\label{boro226c}
\Delta_{_{\!J}}(v_{+}^{})\!\!&=&\!\!v_{+}^{}\otimes
e^{\sigma}+ 1\otimes v_{+}^{}\,,
\\[9pt]\label{boro226d}
\Delta_{_{\!J}}(v_{-}^{})\!\!& =&\!\!v_{-}^{}\otimes e^{-\sigma}+
1\otimes v_{-}^{}+\xi\,h\otimes v_{+}^{}\,e^{-2\sigma}\,,
\end{eqnarray}
where $\sigma=\frac{\xi}{2}E_{+}^{}=\frac{1}{2}\ln(1+\xi e_{+}^{})$.
The coproducts $\Delta_{_{\!J}}(e_{-}^{})$ can be calculated from
(\ref{boro1.12b}) and (\ref{boro226d}). 
We also recall the antipodes  given by the formulae \cite{bor18}
\renewcommand{\theequation}{28\alph{equation}}
\setcounter{equation}{0}
\begin{eqnarray}\label{boro227d}
S_{_{\!J}}(e^{\pm\sigma})\!\!& =\!\!&e^{\mp\sigma}\,,
\\[3pt]
S_{_{\!J}}(e_{+}^{})\!\!& =\!\!&-e_{+}^{}\,e^{-2\sigma}\,,
\\[3pt]
S_{_{\!J}}(h)\!\!& =\!\!&-h\,e^{2\sigma}\,,
\\[3pt]
S_{_{J}}(v_{+}^{})\!\!&=&\!\!-e^{-\sigma}v_{+}^{}\,,
\\[1pt]
S_{_{J}}(v_{-}^{})\!\!&=&\!\!- v_{-}^{}\,e^{\sigma}+
\xi\,h\,v_{+}^{}e^{\sigma}\,.
\end{eqnarray}
It should be added that the coproduct  $\Delta_{_{J}}$ is real
under both involutions (\ref{boro1.13a},\,c), i.e.
\renewcommand{\theequation}{\arabic{equation}}
\setcounter{equation}{28}
\begin{equation}\label{star}
\Delta_{_{J}}(a^\star)=(\Delta_{_{J}}(a))^\star\end{equation} as
well as antipode $S_{_{J}}$ satisfies the consistency condition
(see e.g. \cite{SM}, $\S$ 1.7)
\begin{equation}\label{antypod}
S_{_{J}}((S_{_{J}}(a^\star))^\star)=a
\end{equation}
($a\in U(osp(1|2));\,\star=\dagger,\,\ddagger$) as well as
$\epsilon(a^\star)=\overline{\epsilon(a)}$ is trivially valid. We see
therefore that the formulae (\ref{boro226a}\,-\,e) define two real
quantum $osp(1|2)$ superalgebras subject to a choice of reality
condition (\ref{boro1.13a}) or (\ref{boro1.13b}).

The "super" part $F_{_{\!S}} $ of the twist element
(\ref{boro221}) can be given as solution of Eq. (\ref{boro22.4})
by the following explicite formula
\renewcommand{\theequation}{\arabic{equation}}
\setcounter{equation}{30}
\begin{equation}\label{boro227}
F_{_{\!S}}=1-4\xi\,\frac {v_{+}^{}}{e^{\sigma}+1}\otimes
\frac{v_{+}^{}}{e^{\sigma}+1}\,,
\end{equation}
and obviously
\begin{equation}\label{boro228}
(\epsilon _{_{\!J}}\otimes 1)(F_{_{\!S}})=
(1\otimes\epsilon_{_{\!J}})(F_{_{\!S}})=1\,,
\end{equation}
where $\epsilon_{_{\!J}}(\sigma)=\epsilon_{_{\!J}}(v_{\pm})=0$,
$\epsilon_{_{\!J}}(1)=1$. Further one can rewrite (\ref{boro227}) as follows
\begin{equation}\label{boro229}
F_{_{\!S}}=1-\xi\,\frac{v_{+}e^{-\frac{1}{2}\sigma}}
{\cosh\,\frac{1}{2}\sigma}\otimes \frac{v_+ \, e^{-\frac{1}{2}\sigma}}
{\cosh\,\frac{1}{2}\sigma}\,.
\end{equation}
It can be also shown that the element
\begin{equation}\label{boro2210}
F_{_{\!S}}^{-1} =
\frac{\cosh\frac{1}{2}\sigma\otimes\cosh\frac{1}{2}\sigma
+\xi\,v_{+}\,e^{-\frac{1}{2}\sigma}\otimes
v_+\,e^{-\frac{1}{2}\sigma}}
{\cosh\frac{1}{2}\Delta_{_{\!J}}(\sigma)}\,
\end{equation}
is inverse to $F_{_{\!S}}$, i.e.
$F^{-1}_{_{\!S}}F_{_{\!S}}^{}=F_{_{\!S}}^{}F^{-1}_{_{\!S}}=1$,
where $\Delta_{_{\!J}}(\sigma)=\sigma\otimes1+1\otimes\sigma$.

Using the formula
$\Delta_{_{SJ}}^{}(a)=F_{_{\!S}}^{}\Delta_{_{J}}^{}(a)F^{-1}_{_{\!S}}$,
which follows from (\ref{boro221}), and (\ref{boro226a}-e), (\ref{boro227}),
(\ref{boro2210}) one can calculate explicitly the super-Jordanian twisted
coproducts for $osp(1|2)$. One gets after quite involved calculations that
\renewcommand{\theequation}{35\alph{equation}}
\setcounter{equation}{0}
\begin{eqnarray}\label{boro2.11b}
\Delta_{_{SJ}}(h)\!\!&=&\!\!h\otimes e^{-2\sigma}\!\!+1\otimes h+
\xi v_{+}^{}e^{-\sigma}\!\otimes v_{+}^{}e^{-2\sigma}
-\displaystyle{\frac{1}{4}}\bigl(e^{-\sigma}\!\!-1\bigr)\!
\otimes\!\bigl(e^{-\sigma}\!-1\bigr)e^{-\sigma}\!,\phantom{aaaaaaa}
\\[9pt]\label{boro2.11c}
\Delta_{_{SJ}}(v_{+}^{})\!\!&=\!\!&v_{+}^{}\otimes1+
e^{\sigma}\!\otimes v_{+}^{}\,,
\end{eqnarray}\vskip-18pt
\begin{eqnarray}\label{boro2.11d}
\begin{array}{r}
\Delta_{_{SJ}}(v_{-}^{})=v_{-}^{}\otimes e^{-\sigma}\!+ 1\otimes
v_{-}^{}\!+\displaystyle{\frac{\xi}{4}}\biggl\{\!\Bigl(
\bigl\{h,e^{\sigma}\bigr\}\otimes v_{+}^{}e^{-2\sigma}-
\{h,v_{+}^{}\}\otimes(e^{\sigma}-1)e^{-2\sigma}+
\\[9pt]
+2\,v_+{}^{}\!\otimes h-\Bigl\{h,\displaystyle{
\frac{v_{+}e^{\sigma}}{e^{\sigma}\!+1}}\Bigr\}\!\otimes\!(e^{\sigma}\!-1)
e^{-\sigma}\!+(e^{\sigma}\!-1)\!\otimes\!\Bigl\{h,\displaystyle{\frac{v_{+}}
{e^{\sigma}\!+1}}\Bigr\}\Bigr),\displaystyle{\frac{1}{e^{\sigma}\!\otimes
e^{\sigma}\!+1}}\biggr\}+
\\[14pt]
+\displaystyle{\frac{\xi}{4}}\Bigl(\displaystyle{\frac{v_{+}e^{\sigma}}
{e^{\sigma}\!+1}}\otimes(e^{\sigma}\!\!-1)e^{\sigma}\!+({e^{\sigma}\!\!-1})
e^{\sigma}\otimes\displaystyle{\frac{v_{+}}{e^{\sigma}\!\!+1}}-
v_{+}^{}(e^{\sigma}\!\!-1)\!\otimes(e^{\sigma}\!\!-1)\Bigl)\displaystyle{
\frac{e^{-\sigma}\!\otimes e^{-2\sigma}}{e^{\sigma}\!\otimes e^{\sigma}\!+1}}\,,
\end{array}
\end{eqnarray}
and $\Delta_{_{SJ}}(e_{-})=-4(\Delta_{_{SJ}}(v_{-}^{}))^2$,
where we use the denotation $\{a,b\}:=ab+ba$. It can be added that
the coproduct relations (\ref{boro226a},b) remain still valid.

The formula (\ref{boro2.11c}) and an analog of (\ref{boro2.11b})
(see (\ref{boro2.11'b})) were given by Kulish \cite{bor19}; due to
the explicit knowledge of the super-Jordanian twisting two-tensor
we calculated also the coproduct (\ref{boro2.11d}).

The formulae for the antipode $S_{_{SJ}}$ look as follows:
\renewcommand{\theequation}{36\alph{equation}}
\setcounter{equation}{0}
\begin{eqnarray}\label{boro331}
S_{_{\!SJ}}(h)\!\!&
=\!\!&-h\,e^{2\sigma}+\frac{1}{2}(e^{\sigma}-1)\,,
\\[3pt]
S_{_{SJ}}(v_{+}^{})\!\!&=&\!\!-e^{-\sigma}v_{+}^{}\,,
\\[1pt]
S_{_{SJ}}(v_{-}^{})\!\!&=&\!\!- v_{-}^{}\,e^{\sigma}+
\xi\,h\,v_{+}^{}e^{\sigma}-\frac{1}{2}\,\xi\,v_{+}^{}\frac{e^{\sigma}}{e^{\sigma}+1}\,.
\end{eqnarray}
Supplementing with $\epsilon_{_{\!SJ}}(v_{\pm})=\epsilon_{_{\!SJ}}(h)=0$ we
obtain the complete set of formulae describing the super-Jordanian deformation
$U_\xi (osp(1|2))$ as non-cocommutative Hopf algebra.

It is easy to see that the formulae
(\ref{boro2.11b})--(\ref{boro2.11d}) do not satisfy neither the
reality conditions  (\ref{boro1.13a}) nor (\ref{boro1.13b}), i.e.
$(\Delta_{_{SJ}}(x))^\star\neq\Delta_{_{SJ}}(x^\star)$ for some
element $x\in U_{\xi}(osp(1|2))$, where $\star=\dag$ or $\ddag$.
However, if we require that the super-Jordanian coproduct
$\tilde{\Delta}_{_{SJ}}:=\tilde{F}_{_{S}}^{}
\Delta_{_{J}}\tilde{F}_{_{S}}^{-1}$ satisfy the $\star$-reality
condition it is necessary and sufficient to assume the following
unitarity condition for the twisting two-tensor
$\tilde{F}_{_{S}}^{}$:
\renewcommand{\theequation}{\arabic{equation}}
\setcounter{equation}{36}
\begin{equation}\label{boro332}
\tilde{F}_{_{S}}^{*}=\tilde{F}_{_{S}}^{-1}\,.
\end{equation}
The condition (\ref{boro332}) can be achieved because the twisting
element (\ref{boro221}) is not unique -- without modifying
corresponding $r$-matrix (11b) it can be multiplied by suitable
multiplicative factor $\Phi\in U(osp(1|2))\otimes U(osp(1|2))$
\begin{equation}
\Phi=\frac{f(e^\sigma)\otimes f(e^\sigma)}{f(e^\sigma
\otimes e^\sigma)}\,.
\end{equation}
provided that $f(1)=1$.
It can be shown that modified twisting element
\begin{equation}\label{modi2}
\tilde{F}_{_{\!S}}=\Phi\,F_{_{\!S}}\,,
\end{equation}
satisfies the relations (\ref{boro2.2}) and (\ref{boro2.3}). Different choices
of $f$ in the multiplicative factor $\Phi$ describe non\-uniqueness (up to Hopf
automorphism \cite{bor9}) of the twist quantization of the classical $r$-matrix
(11b). In particular, we can choose the element $f$ such that the twisting
two-tensor will satisfy the reality condition (\ref{boro332}). Indeed,
choosing in (\ref{modi2})
$f(\sigma)=\sqrt{\frac{1}{2}(e^{\sigma}+1)}$ one obtains the
formula for the super-Jordanian two-tensor
\begin{equation}\label{boro20old}
\tilde{F}_{_{\!S}}=\Phi\,\Big(1-4\xi\,\frac {v_{+}^{}}{e^{\sigma}+1}\otimes
\frac{v_{+}^{}}{e^{\sigma}+1}\Big)
\end{equation}
with
\begin{equation}\label{boro20older}
\Phi=\sqrt{\frac{(e^\sigma+1)\otimes(e^\sigma+1)}
{2(e^\sigma\!\otimes e^\sigma+1)}}\,,
\end{equation}
which satisfies with respect to the Hermitean  (\ref{boro1.13a})
and super-Hermitean conjugation (\ref{boro1.13b}) the following
unitarity condition
\begin{equation}\label{boro332'}
\tilde{F}_{_{S}}^{\star}=\tilde{F}_{_{S}}^{-1}\,\quad {\rm
for}\,\, \star=\dag\,\,{\rm or}\,\,\ddag\,,
\end{equation}
provided that the parameter $\xi$ is purely imaginary,
$\xi^{*}:=\bar\xi=-\xi$. 
Such choice will modify the coproduct,
$\tilde{\Delta}_{_{SJ}}=\Phi\Delta_{_{SJ}}\Phi^{-1}$, and we
obtain
\renewcommand{\theequation}{43\alph{equation}}
\setcounter{equation}{0}
\begin{eqnarray}
\label{boro2.11'b}
\tilde{\Delta}_{_{SJ}}(h)\!\!&=&\!\!h\otimes e^{-2\sigma}\!\!+1\otimes h+
\xi v_{+}^{}e^{-\sigma}\!\otimes v_{+}^{}e^{-2\sigma}\,,
\\[9pt]\label{boro2.11'c}
\tilde{\Delta}_{_{SJ}}(v_{+}^{})\!\!&=\!\!&v_{+}^{}\otimes1+
e^{\sigma}\!\otimes v_{+}^{}\,,
\end{eqnarray}\vskip-17pt
\begin{eqnarray}\label{boro2.11'd}
\begin{array}{r}
\tilde\Delta_{_{SJ}}(v_{-}^{})=v_{-}^{}\otimes e^{-\sigma}\!+
1\otimes v_{-}^{}\!+\displaystyle{\frac{\xi}{4}}\biggl\{\!\Bigl(
\bigl\{h,e^{\sigma}\bigr\}\otimes v_{+}^{}e^{-2\sigma}-
\{h,v_{+}^{}\}\otimes(e^{\sigma}-1)e^{-2\sigma}+
\\[12pt]
+2\,v_+{}^{}\!\otimes h-\Bigl\{h,\displaystyle{
\frac{v_{+}e^{\sigma}}{e^{\sigma}\!+1}}\Bigr\}\!\otimes\!(e^{\sigma}\!-1)
e^{-\sigma}\!+(e^{\sigma}\!-1)\!\otimes\!\Bigl\{h,\displaystyle{\frac{v_{+}}
{e^{\sigma}\!+1}}\Bigr\}\Bigr),\displaystyle{\frac{1}{e^{\sigma}\!\otimes
e^{\sigma}\!+1}}\biggr\}\,.
\end{array}
\end{eqnarray}
The formulae for the antipode $\tilde{S}_{_{SJ}}$ look as follows:
\renewcommand{\theequation}{44\alph{equation}}
\setcounter{equation}{0}
\begin{eqnarray}\label{boro331'}
\tilde{S}_{_{\!SJ}}(h)\!\!&
=\!\!&-h\,e^{2\sigma}+\frac{1}{4}(e^{2\sigma}-1)\,,
\\[3pt]
\tilde{S}_{_{SJ}}(v_{+}^{})\!\!&=&\!\!-e^{-\sigma}v_{+}^{}\,,
\\[1pt]
\tilde{S}_{_{SJ}}(v_{-}^{})\!\!&=&\!\!- v_{-}^{}\,e^{\sigma}+
\xi\,h\,v_{+}^{}e^{\sigma}-\frac{\xi}{4}\,v_{+}^{}e^{\sigma}\,.
\end{eqnarray}
It is easy to see that the formulae (\ref{boro2.11'b}--c) satisfy
the reality condition
$(\tilde{\Delta}_{_{SJ}}(a))^\star=\tilde{\Delta}_{_{SJ}}(a^\star)$
for $\star=\dag,\,\ddag$ and any $a\in osp(1|2))$ and the
antipodes (\ref{boro331'}--c) satisfy respectively the condition
(\ref{antypod}). Subequently, we can state that the relations
(\ref{boro2.11'b}--c) and (\ref{boro331'}--c) describe two real
quantum $osp(1|2)$ Hopf superalgebras.

 The universal $R$-matrix has the form
\renewcommand{\theequation}{\arabic{equation}}
\setcounter{equation}{44}
\begin{equation}\label{boro32}
\tilde R_{_{\!SJ}}^{}=\tilde{F}^{21}_{_{\!S}}
R_{_{\!J}}^{}\tilde{F}^{-1}_{_{\!S}}\,,
\end{equation}
where
$\tilde{F}^{-1}_{_{\!S}}=\tilde{F}^{21}_{_{\!S}}(\xi)=\tilde{F}_{_{\!S}}(-\xi)$
and
\begin{equation}\label{boro33}
R_{_{\!J}}=F^{21}_{_{\!J}}F^{-1}_{_{\!J}}=e^{2\sigma \otimes h}
e^{-2h\otimes \sigma}~.
\end{equation}
If $\xi$ is purely imaginary one can show that the universal
$R$-matrices $R_{_{\!J}}$ and $\tilde R_{_{\!SJ}}$ are unitary or
superunitary, what depends on the choice of reality condition and
is inherited from analogous properties of $F_{_{\!J}}$ and $\tilde
F_{_{\!SJ}}$.

The twist deformations described by the formulae
(\ref{boro2.11b}-c) and (\ref{boro2.11'b}-c) locates whole
deformation in coalgebra sector. In order to distribute "more
evenly" the deformation in algebraic and coalgebraic sector one
should introduce the suitable deformation map from classical to
deformed $osp(1|2)$ quantum superalgebra basis. It should be added
that recently an interesting deformed basis in the algebraic
sector of $U_\xi(osp(1|2))$ was proposed \cite{ACS}.
Unfortunately, these authors did not find neither the explicit
formula for two-tensor $F_{_S}$ nor for the universal $R$-matrix.

\section{Quantum $U_{\xi}(osp(1|2))$ and Deformed $D=1$ Superconformal
Mechanics}

The $D=1$ conformal algebra (7) can be extended to the $D=1$
simple superconformal algebra by adding two Hermitean (real) odd
supercharges $Q$ and $S$, describing the "supersymmetric roots" of
the momenta $P$ and conformal momenta $K$ (see e.g.
\cite{AP,bor20,bor21,bor22}). In order to have a standard
description of supercharges in Hilbert space we shall use the
reality condition (\ref{boro1.13a}) with  the antilinear
antiinvolutive mapping $\dagger$ satisfying properties
(\ref{boro1.13a'}).

{}From the reality condition (\ref{boro1.13a}) follow the
following definitions of real supercharges
\begin{eqnarray}\label{supercharge}
Q=\sqrt{-i}\,v_{+}^{}\,,\qquad
S=\sqrt{-i}\,v_{-}^{}
\end{eqnarray}
and from (\ref{boro1.12a}-b) one gets
\renewcommand{\theequation}{48\alph{equation}}
\setcounter{equation}{0}
\begin{equation}\label{boro3.2}
\{Q,\,Q\}=\frac{1}{2}\,{\cal H}\,,\qquad\{S,\,S\}=\frac{1}{2}\,{\cal K}\,,
\qquad\{S,\,Q\}=\frac{1}{2}\,{\cal D}\,.
\end{equation}
The $sp(2;\mathbb R)\simeq o(2,1)$ covariance relations for the supercharges
$Q,S$ look as follows
\begin{equation}\label{boro3.3}
\begin{array}{rcccl}
[{\cal H},\,Q]\!\!& =&\!\!0\,, \qquad\quad\;\;[{\cal H},\,S]\!\!&=&\!\!-i\,Q\,,
\\[5pt]
[{\cal K},\,Q]\!\!&=&\!\!i\,S\,,\qquad\quad [{\cal K},\,S]\!\!&=&\!\!0\,,
\\[5pt]
[{\cal D},\,Q]\!\!&=&\!\!\displaystyle{\frac{i}{2}}\,Q\,,\qquad\;\;
[{\cal D},\,S]\!\!&=&\!\!-\displaystyle{\frac{i}{2}}\,S \, .
\end{array}
\end{equation}
The relations (\ref{boro1.7}) and (\ref{boro3.2})--(\ref{boro3.3})
after taking into consideration the relation (\ref{boro1.8})
and(\ref{supercharge}) can be identified with $osp(1,2)$ algebra.

The algebra (7) can be realized in two-dimensional phase space, generated
by one pair of phase space variables $(x,p)$ satisfying Heisenberg relations
$[x,p]=i$ (we put $\hbar=1$). One can assume (see e.g. \cite{bor23}):
\renewcommand{\theequation}{\arabic{equation}}
\setcounter{equation}{48}
\begin{equation}\label{boro3.4}
{\cal H}=\frac{p^2}{2m}\,,\qquad{\cal D}=\frac{1}{4}\,(px + xp)\,,
\qquad{\cal K}=\frac{1}{2}\,mx^2\,.
\end{equation}
Adding one real fermionic variable satisfying the anticommutation relation
\begin{equation}\label{boro3.5}
\{\psi,\,\psi\}=1\,,
\end{equation}
one can realize the superalgebra (\ref{boro3.2},b) if we supplement the
generators (\ref{boro3.4}) by the following  odd ones
\begin{equation}\label{boro3.6}
Q=\frac{p\cdot\psi}{\sqrt{4m}}\,,\qquad S=\sqrt{\frac{m}{4}}\,x\cdot\psi\,.
\end{equation}
The realization (\ref{boro3.4})--(\ref{boro3.6}) can be
generalized to $N$--dimensional superconformal mechanics in curved
target superspace with a  suitable torsion \cite{bor24,bor25},
with the generators described by 2N bosonic phase space variables
($x_i , p_i$) ($i=1,\ldots N$) and N real vectorial fermionic
variables $\psi_i$, satisfying $N$--dimensional Clifford algebra.

We would like to point out that quantum $\kappa$-deformed
$osp(1|2)$ superalgebra, obtained via twist quantization
technique, can be applied to the deformation of supersymmetric
conformal mechanics. If we use the classical $osp(1|2)$ basis,
satisfying the relations (\ref{boro3.2}), (\ref{boro3.3}), one can
use coproduct of energy operator ${\cal H}$ in order to describe
the two--particle interactions \footnote{For $D=4$ relativistic
case an attempt in this direction leading to analogous formula was
proposed in \cite{bor26}}. Using the formulae (\ref{boro1.12b}),
(\ref{supercharge}), (\ref{boro2.11'c}) and the property of graded
tensor product one gets (recall that $i\,{\cal
H}=e_{+}^{}=4v_{+}^2=4i\,Q^2$ and we put
$\xi=\frac{1}{2i\,\kappa}$)
\begin{equation}\label{boro3.7}
\Delta({\cal H})=4\,(\Delta(Q))^2={\cal H}\otimes1 +
(1+\frac{1}{2\kappa}{\cal H})\otimes{\cal H}\,,
\end{equation}
i.e. we obtain the energy of two--particle system described by the
formula
\begin{equation}\label{boro3.8}
{\cal E}_{1+2}^{}={\cal E}_{1}^{}+{\cal E}_{2}^{}+
\frac{1}{2\kappa}{\cal E}_{1}^{}\cdot{\cal E}_{2}^{}\,.
\end{equation}
We see that the deformation parameter $ \kappa$ describes a
geometric mass parameter and its inverse $\frac{1}{\kappa}$ can be
interpreted as coupling constant in superconformal two--particle
dynamics. Surprisingly, the formulae (\ref{boro3.7}),
(\ref{boro3.8}) are the same as for the well-known bosonic
Jordanian deformation of $sl(2)\simeq o(2,1)$, given by the twist
(\ref{boro1.4}); the supersymmetric corrections depending on the
supercharges (\ref{supercharge}) will be however presented in
coproducts  of ${\cal D}$ and ${\cal K}$.

In this section we applied the quantum deformation of
$osp(1|2,\mathbb R)$ with the reality condition (\ref{boro1.13a}).
It is interesting to find an application of quantum deformation of
$osp(1|2,\mathbb R)$ with the reality condition (\ref{boro1.13b})
defined by the graded antiinvolution $\ddagger$ which, as we
expect, can be employed in the superspace formulation of dynamical
models.

\section{Outlook}

The super--Jordanian twist and super--Jordanian deformation of
$osp(1|2)$ should play a basic role in the description of twist
quantizations of superalgebras. The role of $osp(1|2)$ in the
theory of superalgebras is analogous to the role played by $sl(2)$
in the theory of Lie groups and Lie algebras. In particular any
superextension of the twist quantization techniques of classical
Lie algebras (see e.g. \cite{bor27,bor28}) requires the twist
element (\ref{boro1.4}) as its basic building block. We expect
that the results of this paper can help substantially for  the
construction of extended twist elements for arbitrary
orthosymplectic algebra $osp(N|2M)$, which for $M=1$ and arbitrary
$N$ should describe  $N$--extended supersymmetric conformal
mechanics, and for $M=2$ has physical interpretation as describing
four--dimensional anti--de--Sitter supersymmetries.

Also it should be recalled here that the twist element (4) has been used for
deformations of quantum infinite--dimensional algebras, e.g. the
Yangians  \cite{bor29, bor29'}. Having an explicite form of
super--Jordanian twist (\ref{boro221}) it is a matter of standard
calculation to obtain the twisted form of $osp(2|1)$
super--Yangian\footnote{The $osp(1|2)$ super--Yangian has been
described recently by one of the authors \cite{bor30}.}. Other
possible application of our deformation of $osp(1|2)$ are
integrable vertex models (see e.g. \cite{bor31}) and $osp(1|2)$
Gaudin lattice models \cite{bor32,bor33}. Also the knowledge of
super-Jordanian twist (14) permits to calculate the Clebsch-Gordon
coefficients of $\kappa$-deformed $osp(1|2)$ by using the
technique of projection operators proposed in \cite{bor13}.

\subsection*{Acknowledgments}
The paper has been supported by KBN grant 5PO3B05620 (JL) and the
grants RFBR-02-01-00668, INTAS OPEN 00-00055 and CRDF
RMI-2334-MD-02 (VNT). A.B. would like to thank the Catedra de
Investigaciones "Algebra y Logica" en UNAM, FESC, M\'{e}xico for
financial support. V.N. Tolstoy would like to thank Institute for
Theoretical Physics University of Wroc{\l}aw for hospitality. One
of the authors (JL) wishes to acknowledge the financial support of
Valencia University and of grant BFM2002-03681.


\begin{thebibliography}{99}

\bibitem{PK} P.P. Kulish and N.Yu. Reshetikhin,
Journ. Sov. Math. {\bf 23}, 2435 (1983).

\bibitem{bor1} V.G. Drinfeld, "Quantum Groups", in Proc. XX-th Int.
Congress of Math. (Berkeley, USA, 1986), p. 798.

\bibitem{bor2} M. Jimbo, Lett. Math. Phys. {\bf 10}, 63 (1985).

\bibitem{bor3} L.D. Faddeev, N.Yu. Reshetikhin and L.A. Takhtadjan,
Algebra i Analiz, {\bf 1}, 178 (1989).

\bibitem{bor4} M. Dubois-Violette and G. Launer, Phys. Lett. {\bf
B245}, 175 (1990).

\bibitem{bor6} O.V. Ogievetsky, Suppl. Rendic. Cir. Math. Palermo,
Serie II, No 37, p. 185 (1993); preprint MPI-Ph/92-99 (1992).

\bibitem{bor7} C. Ohn, Lett. Math. Phys. {\bf 25}, 85 (1992).

\bibitem{bor8} V.G. Drinfeld, DAN USSR {\bf 273}, 531 (1983).

\bibitem{bor9} V.G. Drinfeld, Leningrad. Math. Journ. {\bf 1}, 1415
(1990).

\bibitem{VL}  V. Lyubashenko, in "Quantum Groups", ed. P.P.
Kulish, Springer Lecture Notes in Math. N.1510, p. 67 (1992).

\bibitem{bor11} J. Lukierski, P. Minnaert and M. Mozrzymas, Phys. Lett.
{\bf B371}, 215 (1996).

\bibitem{bor12} M. Scheunert, W. Nahm and V. Rittenberg, J. Math. Phys.
{\bf 18}, 146 (1977).

\bibitem{bor13} F.A. Berezin and V.N. Tolstoy, Comm. Math. Phys. {\bf 78},
409 (1981).

\bibitem{bor14} C. Juszczak and J. Sobczyk, Czech. J. Phys. {\bf 48},
1375 (1998).

\bibitem{bor15} P.P. Kulish, Preprint RIMS-G15  (1988).

\bibitem{bor16} P.P. Kulish and N.Yu. Reshetikhin, Lett. Math. Phys.
{\bf 18}, 143 (1989).

\bibitem{SM} S. Majid, "Foundations of Quantum Groups Theory",
Cambridge Univ. Press 1995.


\bibitem{bor18} E. Celeghini and P.P. Kulish, J. Phys. {\bf A31}, L79;
preprint {\tt q-alg/9712024}.

\bibitem{bor19} P.P. Kulish, preprint {\tt math.QA/9806104}.

\bibitem{ACS}  N. Aizawa, R. Chakrabarti and J. Segar,
preprint {\tt math.QA/0301022}.

\bibitem{AP} V.P. Akulov and A.I. Pashnev, Teor. Mat. Fiz. {\bf
56}, 344 (862 English version) (1983).

\bibitem{bor20} S. Fubini and E. Rabinovici, Nucl. Phys. {\bf B245}, 17 (1984).

\bibitem{bor21} P. Claus, M. Derix, R. Kallosh, J. Kumar, P. Townsend and
A. van Proyen, Phys. Rev. Lett. {\bf 81}, 4553 (1998).

\bibitem{bor22} J.A. de Azcarraga, J.M. Izquierdo, J.C. Perez-Bueno and
P.K. Townsend, Phys. Rev. {\bf D59}, 084015 (1999).

\bibitem{bor23} V. de Alfaro, S. Fubini and G. Furlan, Nuovo Cimm. {\bf 34A},
569 (1978).

\bibitem{bor24} J. Michelson and A. Strominger, Comm. Math. Phys. {\bf 213},
1  (2000), {\tt hep-th/9907191}.

\bibitem{bor25} R. Britto-Pacumio, J. Michelson, A. Strominger, and A. Volovich,
JHEP {\bf 9911}, 013 (1999), {\tt hep-th/9911066}.

\bibitem{bor26} J. Lukierski and A. Nowicki, in the Proceedings of Quantum Group
Symposium at "Group21", eds. H-D. Doebner and V.K.Dobrev, Heron Press, Sofia,
1997, p. 173 (see \S\ 2).

\bibitem{bor27} P.P. Kulish, V.D. Lyakhovsky and A.I. Mudrov, J. Math. Phys.
{\bf 40}, 4569 (1999).

\bibitem{bor28} V.D. Lyakhovsky and M.A. del Olmo, Journ. Phys. {\bf A32},
4541 (1999); {\bf A32}, 5343 (1999).

\bibitem{bor29} S.M. Khoroshkin, A.A. Stolin and V.N. Tolstoy, Commun.
Algebra, {\bf 26}, no. 4, 1041 (1998).

\bibitem{bor29'} S.M. Khoroshkin, A.A. Stolin and V.N. Tolstoy,
Yadernaya Fizizika, {\bf 64}, No. 12, 2262 (2001); 
Physics of Atomic Nuclei, {\bf 64}, No. 12, 2173 (2001); 
{\tt arXiv:math.QA/0012207}.

\bibitem{bor30} V.N. Tolstoy, SQS`99, Proc. of Dubna Workshop "Supersymmetries
and Quantum Symmetries", Jul 1999, eds. E. Ivanov et al., Dubna 2000, p. 431.

\bibitem{bor31} H. Saleur, Nucl. Phys. {\bf B336}, 363 (1990).

\bibitem{bor32}  T. Brzezinski and A.J. Mecfarlane, J. Math. Phys. {\bf 35},
3261 (1994).

\bibitem{bor33} P.P. Kulish, N. Manajlovic, J. Math. Phys. {\bf 42}, 4757 (2001).

\end{thebibliography}
\end{document}